\documentstyle[12pt,epsfig,preprint]{aastex}

\tighten

\slugcomment{submitted to ApJ.}
\newcommand{\mpc}{\rm {h^{-1}Mpc }}
\newcommand{\etal}{{\it et al.\ }}

\newcommand{\avg}[1]{\langle{#1}\rangle}

\newcommand{\ltsima}{$\; \buildrel < \over \sim \;$}
\newcommand{\lsim}{\lower.5ex\hbox{\ltsima}}
\newcommand{\gtsima}{$\; \buildrel > \over \sim \;$}
\newcommand{\gsim}{\lower.5ex\hbox{\gtsima}}

\begin{document}

\title{Higher Order Moments of the Angular Distribution of Galaxies 
from Early SDSS Data}

\author{Istv\'an Szapudi\altaffilmark{1},
Joshua A. Frieman\altaffilmark{2,3},
Roman Scoccimarro\altaffilmark{11},
Alexander S. Szalay\altaffilmark{13},
Andrew J. Connolly\altaffilmark{14},
Scott Dodelson\altaffilmark{2,3},
Daniel J. Eisenstein\altaffilmark{2,4,27}, 
James E. Gunn\altaffilmark{5},
David Johnston\altaffilmark{2,3},
Stephen Kent\altaffilmark{3},
Jon Loveday\altaffilmark{8}, 
Avery Meiksin\altaffilmark{26}
Robert C. Nichol\altaffilmark{9}, 
Ryan Scranton\altaffilmark{2,3}, 
Albert Stebbins\altaffilmark{3},
Michael S. Vogeley\altaffilmark{15},
James Annis\altaffilmark{3},
Neta A. Bahcall\altaffilmark{5}
J. Brinkman\altaffilmark{16}
Istv\'an Csabai\altaffilmark{17}
Mamoru Doi\altaffilmark{18},
Masataka Fukugita\altaffilmark{18}, 
\v{Z}eljko Ivezi\'c\altaffilmark{5},
Rita S.J. Kim\altaffilmark{5},
Gillian R. Knapp\altaffilmark{5},
Don Q. Lamb\altaffilmark{2},
Brian C. Lee\altaffilmark{3},
Robert H. Lupton\altaffilmark{5}, 
Timothy A. McKay\altaffilmark{23},
Jeff Munn\altaffilmark{24},
John Peoples\altaffilmark{3},
Jeff Pier\altaffilmark{24}, 
Constance Rockosi\altaffilmark{2},
David Schlegel\altaffilmark{5}, 
Christopher Stoughton\altaffilmark{3},
Douglas L. Tucker\altaffilmark{3},
Brian Yanny\altaffilmark{3}, 
Donald G. York\altaffilmark{2,25}, 
for the SDSS Collaboration
}
 
\altaffiltext{1}{Institute for Astronomy, University of Hawaii, 2680
Woodlawn Drive, Honolulu, HI 96822, USA} 
\altaffiltext{2}{Astronomy and Astrophysics Department, University of
Chicago, Chicago, IL 60637, USA} 
\altaffiltext{3}{Fermi National Accelerator Laboratory, P.O. Box 500,
Batavia, IL 60510, USA} 
\altaffiltext{4} {Steward Observatory, University of Arizona, 933 N.\
Cherry Ave ., Tucson, AZ 85721} 
\altaffiltext{5}{Princeton University Observatory, Princeton, NJ
08544, USA} 
\altaffiltext{6}{Department of Physics, Columbia University, New York,
NY 10027, USA} 
\altaffiltext{7}{Department of Physics, University of Pennsylvania,
Philadelphia, PA 19101, USA} 
\altaffiltext{8}{Sussex Astronomy Centre, University of Sussex,
Falmer, Brighton BN1 9QJ, UK} 
\altaffiltext{9}{Department of Physics, 5000 Forbes Avenue, Carnegie
Mellon University, Pittsburgh, PA 15213, USA} 
\altaffiltext{10}{Space Telescope Science Institute, Baltimore, MD
21218, USA} 
\altaffiltext{11}{Department of Physics, New York University, 4
Washington Place, New York, NY 10003} 
\altaffiltext{13}{Department of Physics and Astronomy, The Johns
Hopkins University, Baltimore, MD 21218, USA}
\altaffiltext{14}{University of Pittsburgh, Department of Physics and
Astronomy, Pittsburgh, PA 15260, USA} 
\altaffiltext{15}{Department of Physics, Drexel University,
Philadelphia, PA 19104, USA} 
\altaffiltext{16}{Apache Point Observatory, P.O. Box 59, Sunspot, NM
88349-0059} 
\altaffiltext{17}{Department of Physics of Complex
Systems,E\"{o}tv\"{o}s University, Budapest, Hungary, H-1088}
\altaffiltext{18}{University of Tokyo, Institute of Astronomy and
Research Center of the Early Universe, School of Science}
\altaffiltext{19}{University of Tokyo, Institute for Cosmic Ray
Research, Kashiwa, 2778582} 
\altaffiltext{20}{US Naval Observatory, 3450 Massachusetts Ave., NW,
Washington, DC 20392-5420} 
\altaffiltext{25}{Remote Sensing Division, Naval Research Laboratory,
45555 Overlook Ave.  SW, Washington DC 20375}
\altaffiltext{22}{Astronomical Institute, Tohoku University, Aoba,
Sendai 980-85 78, Japan} 
\altaffiltext{23}{University of Michigan, Department of Physics, 500
East University, Ann Arbor, MI, 48109} 
\altaffiltext{24}{US Naval Observatory, Flagstaff Station, P.O.\ Box
1149, Flags taff, AZ 86002-1149} 
\altaffiltext{25}{Enrico Fermi Institute, 5640 South Ellis Avenue,
Chicago, 60637} 
\altaffiltext{26}{University of Edinburgh, Royal Observatory, Blackford Hill,
Edinburgh EH9 3HJ, UK} 

\begin{abstract}

We present initial results for counts in cells statistics of the
angular distribution of galaxies in early data from the Sloan Digital
Sky Survey (SDSS).  We analyze a rectangular stripe $2.5^\circ$ wide,
covering approximately 160 sq.  degrees, containing over $10^6$
galaxies in the apparent magnitude range $18 < r^\prime < 22$, with
areas of bad seeing, contamination from bright stars, ghosts, and high
galactic extinction masked out.  This survey region, which forms part
of the SDSS Early Data Release, is the same as that for which
two-point angular clustering statistics have recently been computed. 
The third and fourth moments of the cell counts, $s_3$ (skewness) and 
$s_4$ (kurtosis),
constitute the most accurate measurements to date of these quantities
(for $r^\prime < 21$) over angular scales $0.015^\circ-0.3^\circ$. 
They display the approximate hierarchical scaling expected from
non-linear structure formation models and are in reasonable agreement
with the predictions of $\Lambda$-dominated cold dark matter models with
galaxy biasing that suppresses higher order correlations at small scales.  
The  results are in general consistent with previous measurements in the
APM, EDSGC, and Deeprange surveys. These results suggest that the SDSS 
imaging data are free of systematics to a high degree and 
will therefore enable determination of the 
skewness and kurtosis to  1\% and less then 10\%,  
as predicted by Colombi, Szapudi, \& Szalay (1998).

\end{abstract}


\setcounter{footnote}{0}

\section{Introduction} 

In contemporary models of structure formation, initially small density
fluctuations are amplified by gravitational instability to form
non-linear structures.  Within such structures, the baryons undergo
additional non-gravitational processing (gas dissipation, star
formation, etc), leading eventually to luminous galaxies.  In
principle, therefore, the spatial distribution of galaxies encodes a
wealth of information about the initial conditions for structure
formation, gravitational evolution, the relationship (bias) 
between galaxies and
mass, and the processes involved in galaxy formation.  To extract this
information, the statistical properties of the galaxy distribution
must be measured and compared with the predictions of structure
formation models.

To date, two-point statistics of the cosmic microwave background
and the large scale structure, i.e., the two-point correlation
function and its Fourier transform the power spectrum, have been the
primary means of testing structure formation models.  This is
appropriate, since two-point statistics provide the lowest-order
measures of departure from homogeneity; moreover, for Gaussian
fluctuations, they provide a complete statistical description of the
density field.  However, to precisely probe structure formation
models, it is necessary to go beyond second-order information, for: i)
the observed galaxy distribution {\em is} non-Gaussian and, in
particular, its web-like spatial coherence is not captured by
two-point information; ii) the dissipative processes of galaxy
formation in general imply that the galaxy distribution differs from
the underlying dark matter field. This latter bias between galaxies and mass
(Kaiser 1984; Davis \etal 1985; Bardeen \etal 1986) is only partly
constrained by two-point statistics.

Given these limitations of the two-point function, in this paper we
consider higher order statistics of the galaxy distribution.  The
theory of higher order statistics is well-developed (e.g., Peebles
1980; Fry 1984; Bernardeau 1992, 1994; Juszkiewicz, Bouchet \& Colombi
1993; Bouchet \etal 1993; Colombi, Bouchet \& Schaeffer 1995; Szapudi
\etal 1998; Colombi \etal 2000; Szapudi \etal 2000).  In particular,
it has been shown that higher order moments of the galaxy distribution
provide important constraints on non-Gaussianity in the initial
conditions (Fry \& Scherrer 1994; Jaffe 1994; Chodorowski \& Bouchet 1996;
Gazta\~naga \& M\"ah\"onen 1996l; Frieman \& Gazta\~naga 1999;
Scoccimarro 2000; Feldman \etal 2001; 
Durrer \etal 2000)
and on the bias (Fry \& Gazta\~naga 1993; Frieman \& 
Gazta\~naga 1994; Gazta\~naga \& Frieman 1994, Fry 1994;
Juszkiewicz \etal 1995;  Matarrese, Verde and Heavens 1997;
Frieman \& Gazta\~naga 1999;
Szapudi 1999; Scoccimarro \etal 2000; Feldman \etal 
2001).

Here, we measure the $s_p$ parameters (which characterize the
connected moments of counts in cells, hereafter CIC) of the angular
galaxy distribution in early SDSS data; these can be viewed as
normalized area averages of the projected (angular) $p$-point
correlation functions.  We use the same data set for which
second-order statistics were recently reported by Connolly \etal 2001,
Dodelson \etal 2001, Scranton \etal 2001, Szalay \etal 2001, and
Tegmark \etal 2001.  Direct estimation of the angular and
redshift-space three-point correlation functions in early SDSS data
will be presented elsewhere.

Previous measurements of angular galaxy CIC  include those
obtained for the Automated Plate Measuring (APM, 
Gazta\~naga 1994, Szapudi \etal 1995) and Edinburgh Durham
Souther Galaxy Catalog (EDSGC, Szapudi, Meiksin, \& Nichol 1996) 
surveys.  Both have a wide area
covering $\sim 1$ steradian to $b_J \sim 20$ and constructed from
digitized UK Schmidt plates. Similar measurements were performed
in the Deeprange
catalog (Postman \etal 1998, Szapudi \etal 2000), based on a deeper
($I < 23.5$), small-area (16 sq.  deg.)  CCD imaging survey.  The SDSS
sample considered here is intermediate between these two extremes in
both area ($\sim 160$ sq.  deg) and depth ($r < 22$).  We note
that all of these datasets contain of order $10^6$ galaxies with
measured angular positions and apparent magnitudes.  While the SDSS
dataset is based on only two nights of early imaging data and
constitutes less than 2\% of the eventual survey area, the high
quality of the data has enabled us to extract statistical measurements
with smaller errors than those for the earlier catalogs; the SDSS
results therefore provide an important consistency check on and
interpolation between the previous measurements. According
to Colombi, Szapudi \& Szalay (1998) the full SDSS will pinpoint
the skewness and kurtosis with smaller then 1\% and 10\% errors,
respectively,  in a large dynamic range up to $10\mpc$.

The next Section briefly describes the SDSS data set used, while
Section 3 presents an outline of the method used for analysis. 
Section 4 contains the measurements and various tests for
systematic effects, while Section 5 is devoted
to comparisons with measurements in
other surveys. We present our conclusions in Section 6.

\section{The Sloan Digital Sky Survey Early Data Set}

The Sloan Digital Sky Survey is a wide-field photometric and
spectroscopic survey being undertaken by the Astrophysical Research
Consortium at Apache Point Observatory in New Mexico (York \etal
2000).  The completed survey will cover approximately 10,000 square
degrees.  CCD imaging with the SDSS camera (Gunn \etal 1998) will
include $10^8$ galaxies in five passbands ($u, g, r, 
i,$ and
$z$; see Fukugita \etal 1996) to an approximate detection limit of
$r = 23$ at signal-to-noise $S/N =5$ (strictly true for point
sources).

In this paper, we focus on a small section of imaging data that was
taken on the nights of March 20 and 21, 1999, during commissioning of
the survey telescope; these data are designated Runs 752 and 756 and 
are in the Northern Equatorial Stripe of the survey. 
These interleaved scans are centered on the Celestial Equator,
covering a stripe $2.52$ deg wide, with declination $|\delta| <
1.26^\circ$, and approximately 90 deg long, with Right Ascension
ranging from $9^h40^m48^s$ to $15^h45^m12^s$ (J2000).  These data,
designated EDR-P and discussed extensively by Scranton \etal (2001), 
constitute a high-quality subset of the data made publicly
available as part of the SDSS Early Data Release (Stoughton \etal
2001).

The imaging data were reduced by the SDSS photometric image processing
software ({\tt photo}), which detects objects and measures their
apparent magnitudes based on the best-fit PSF-convolved de Vaucouleurs
or exponential model, including an arbitrary scale size and axis
ratio.  All magnitudes were corrected for Galactic extinction using
the reddening maps of Schlegel, Finkbeiner, \& Davis (1998).  For this
dataset, two methods of star-galaxy separation were available.  The
version of the photometric pipeline used to reduce these data makes a
binary decision about whether a given detected object is stellar or
galactic.  For most of the results shown below, we selected all
objects flagged as galaxies in this way by {\tt photo}.  While the
{\tt photo} separation works well at bright magnitudes, a Bayesian
star-galaxy separation method (Scranton \etal 2001), which assigns a
probability to each object of being a star or galaxy, provides a more
precise separation at faint magnitudes.  Below, we also show results
for a sample of objects for which the Bayesian method assigns a very
high probability ($p>0.99$) of their being galaxies.  We find that
these two samples yield consistent results for the higher order
moments.

Observing conditions, particularly the seeing, varied considerably
over the course of these two nights of data taking.  Scranton \etal
(2001) carried out extensive tests of the effects of seeing, galactic
extinction, variable stellar density, etc., on the integrity of the
data.  These tests showed that the data were essentially free of
systematic contamination (as demonstrated by cross-correlation
analysis), provided certain regions were excluded, and we apply the
same masks here.  The resulting masks depend on apparent magnitude: a
``bright'' seeing mask (used for $r < 21$) excludes regions
where the seeing disk is larger than $1.75^{\prime\prime}$, while the
``faint'' seeing mask (used for $21 < r < 22$) excludes regions
for which the seeing exceeds $1.6^{\prime\prime}$.  These masks
exclude roughly 30\% of the data area.  In addition, regions where the
reddening is greater than 0.2 mag in $r$ and small rectangles
around saturated stars and ghost images were also masked out.  The
resulting catalog contains $1.46\times 10^6$ galaxies in the apparent
magnitude range $18 < r < 22$ and covers an area of 160 (140)
square degrees for the bright (faint) samples.  The resulting data
region has a complex geometry (see Figs.  13-14 in Scranton \etal
2001), complicating the CIC analysis, as described below.  Since the
data tests described in Scranton \etal (2001) were based solely on
two-point analyses, the results shown below provide additional
constraints on the data quality: the consistency of the $s_p$
measurements with both previous results and theoretical predictions
indicates that stellar and other contamination of the sample is small.

As in Connolly \etal 2001, Dodelson \etal 2001, Scranton \etal 2001,
Szalay \etal 2001, and Tegmark \etal 2001, we divide the sample into
four (`de-reddened') apparent magnitude slices, $r = 18-19,
19-20, 20-21, 21-22$.  We also consider a ``pseudo APM'' slice, with
$r = 15.9-18.9$, in order to compare with the results of
Gazta\~naga (2001a), as well as a series of slices with $r =
17.9-18.9, 18.9-19.9, 19.9-20.9, 20.9-21.9$ for direct comparison with
the results from the Deeprange survey (Postman \etal 1998, Szapudi
\etal 2000).

\section{The Counts in Cells Method}

The probability distribution of counts in cells (CIC), $P_N(\theta)$, is the
probability that an angular cell of (linear) dimension $\theta$
contains $N$ galaxies.  The factorial moments of this distribution are
defined by $F_k \equiv \sum_N P_N (N)_k$, where $(N)_k =
N(N-1)..(N-k+1)$ is the $k$-th falling factorial of $N$.  The
factorial moments are closely related to the moments of the underlying
continuum random field (which is assumed Poisson-sampled by the galaxies),
$\rho = \avg{N}(1+\delta)$, through $\avg{(1+\delta)^k} =
F_k/\avg{N}^k$ (Szapudi \& Szalay 1993), where angle brackets in the
last relation denote an area average over cells of size $\theta$.  The
factorial moments therefore provide a convenient way to estimate the
angular connected moments, $s_p \equiv \langle\delta^p \rangle_c /
\langle \delta^2 \rangle^{p-1}$, where the subscript $c$ denotes the
connected contribution, and $\avg{\delta^p}_c$ denotes the area
average (over scale $\theta$) of the $p-$point angular correlation
function.  The moments $s_3$ (skewness) and $s_4$ (kurtosis) quantify
the lowest-order deviations of the angular distribution from a
Gaussian.

To measure the $s_p$ amplitudes from CIC via factorial
moments, we must estimate the distribution of CIC in the
survey.  We use the infinite oversampling algorithm of Szapudi (1997),
which eliminates measurement errors due to the finite number of
sampling cells.  From the factorial moments, the recursion relation of
Szapudi \& Szalay (1993) is used to obtain the $s_p$'s.  This
technique is described in more complete detail in Szapudi, Meiksin \&
Nichol (1996) and Szapudi \etal (2001).

The masks described in section 2 represent the most serious practical
problem for CIC estimation, since they generate a complex geometry for
the survey area.  In addition to the seeing masks, there are a large
number of small cut-out holes around bright stars, ghosts, etc.,
yielding about 20,000 separate mask regions.  Since the cut-out holes
are distributed across the entire survey area,
a randomly placed cell larger than about $0.1^\circ$ on a side has a
high probability of intersecting a mask.  In the standard CIC
technique, cells that overlap mask regions are discarded, since one
has no information about the galaxy field in the masked portion of the
cell.  In principle, this would limit our analysis to cells smaller
than $\simeq 0.1^\circ$.

To remedy this situation, we follow Szapudi \etal (2001) and carry out
the CIC analysis by ignoring all masks with area smaller then $0.0021$
square degrees; this leaves only about 300 of the largest masks.
Using only the large masks 
allows us to extend the measurements to cell sizes of order $1^\circ$. 
We test the validity of this prescription by comparing
measurements of the $s_p$ with all the masks and with only the large
masks, for cells smaller than $ 0.1^\circ$.  We find that we
can reliably use just the large masks to make measurements on larger
scales, since the discarded small masks occupy a falling fraction of
the cell area as the cell size increases.

The errors on the $s_p$ measurements shown in the figures were
estimated by using the FORCE (Fortran for Cosmic Errors) package
(Szapudi \& Colombi 1996, Colombi, Szapudi, \& Szalay 1998, Szapudi,
Colombi \& Bernardeau 1999).  This method provides the most accurate
estimation of the errors from the data itself and takes into account
the contributions from the six (eight) point correlations to the
errors on $s_3$ ($s_4$).  It includes error contributions from
finite-volume effects, edge effects, and discreteness.  The
FORCE-estimated errors depend on a number of parameters, 
several of which can be accurately estimated from the data itself (e.g., the 
mean cell occupation number, the angular two-point function averaged 
over the area of a cell, and the effective area of the survey). 
With one exception, 
the remaining parameters are 
obtained from models, e.g., Extended Perturbation Theory 
(Colombi \etal 1997)
is used to estimate $s_6$ and $s_8$; 
as shown by Szapudi, Colombi \& Bernardeau (1999), the $s_p$ errors are
relatively insensitive to variations of these model parameters 
within their presently acceptable ranges.  
The final parameter needed is the estimate for the area-average of the
two-point correlation function over the full survey region, which we
have measured from 100 $\Lambda$CDM realizations (including galaxy
bias) obtained with the {\tt PTHalos} code (Scoccimarro \& Sheth
2001).  These simulations of the galaxy distribution are generated by
using second-order Lagrangian perturbation theory to follow clustering
at large scales and to identify dark matter halos, which are then
replaced by nonlinear dark matter profiles to build the small-scale
correlations.  Galaxy distributions are obtained by sampling the dark
matter profiles with a number of galaxies that depends on the halo
mass $m$, according to $N_{\rm gal}(m) \sim m^{0.8-0.9}$. This 
galaxy-halo scaling leads to an 
approximate match between the measured angular two-point function in 
early SDSS data and the $\Lambda$CDM model; see Scranton
\etal (2001) for a detailed description of this scaling relation and
comparison of the {\tt PTHalos} angular two-point correlation with
measurements in the data.

We also use the {\tt PTHalos} realizations to independently 
estimate $s_p$ errors and
their covariance matrix by averaging over the Monte Carlo pool.  Due
to the significant computational cost required to measure higher-order
moments, however, we have only used this approach for the $r = 
18-19$ and
19-20 magnitude slices, which have the fewest galaxies.  
Directly comparing the FORCE and {\tt PTHalos} errors in this
case for both $s_{3}$ and $s_{4}$, we found that they agree to better
than 30\%, with the FORCE errors being larger (smaller) than the {\tt PTHalos}
errors for angular scales larger (smaller) than 0.05 degrees. 
Given the different assumptions that go into the two
methods, 
this consistency is quite encouraging; while 
the FORCE uses as input the measured higher order
correlations and assumes the hierarchical model ($s_{p} = const.,$ 
independent of $\theta$), the 
{\tt PTHalos} estimate is based on correlations built from 
$\Lambda$CDM models plus a
bias prescription which does not necessarily lead to $s_{p} = const.$
(see, e.g., the solid curves in Fig.~\ref{fig:force} below).  In addition,
the elongated geometry of the survey region and the large number of
masks present an extra challenge for the FORCE method; the agreement 
with the {\tt PTHalos} estimates for the two bright slices indicates 
that the FORCE maintains accuracy under these conditions.

The FORCE error estimation is based on a series expansion: when the
relative errors approach unity, it breaks down.  When this occurs,
the FORCE correctly indicates that the errors are qualitatively large, but
the actual size of the errorbars has no precise statistical meaning in
this regime.  Finally, while the FORCE method is able to calculate the
cosmic bias of the estimators of $s_p$ (Hui \& Gazta\~naga 1999;
Szapudi, Colombi \& Bernardeau 1999), its effects are always negligible in
the regime where the FORCE errors are accurate (i.e., smaller than 
$\sim 100$\%); therefore we do not include a cosmic bias estimate.

It should be borne in mind that error estimation using the FORCE or the {\tt
PTHalos} mock catalogs each represent model-dependent statistical
error calculations.  This is unavoidable, since the errors are
determined in part by galaxy clustering over scales larger than the
survey region, which we cannot directly measure.  The accuracy of the
error estimates is determined by the difference between the assumed
model and the actual distribution.  In addition, these estimates do 
not include possible systematic errors in the data; however, the tests
described in Scranton \etal (2001), although only performed for
two-point statistics, suggest the latter are not dominant over
the scales we consider.  The FORCE error estimates are easily obtained
for faint magnitudes where the {\tt PTHalos} estimates require costly
computational resources due to the relatively large number of
galaxies.  On the other hand, the {\tt PTHalos} method 
can deal trivially with
complex masks and edge effects; in addition, it provides the 
covariance
matrix of the errors, essential for testing models since
measurements at different scales are correlated.

\section{Measurements of Counts in Cells Statistics}

We have carried out a series of measurements of CIC statistics in the
early SDSS angular clustering data.  We first present our principal
results for the normalized connected moments $s_3$ and $s_4$ and their
estimated statistical errors.  To ascertain the reliability of these
results, we have performed a number of auxiliary measurements aimed at
quantifying the possible level of systematic errors.  The most
important of these tests are discussed in Sec.~4.2.

\subsection{Principal Measurements}

\begin{figure*} 
[htb]
\centerline{\hbox{\epsfig{figure=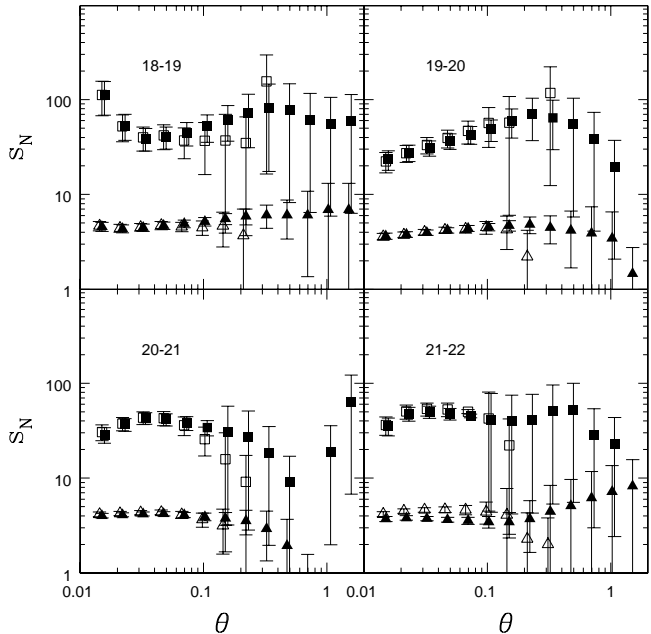,width=9cm}}}
\caption{Results for the angular amplitudes $s_3$ (triangles) and 
$s_4$ 
(squares) in the 
early SDSS data set, in four bins of apparent magnitude ($r$), 
with error estimates from the FORCE.  
Open (closed) symbols denote measurements with all masks (large 
masks only). For clarity, 
the points and errorbars have been shifted slightly 
between these two cases.
For clarity, the largest relative errorbars plotted on the points 
are $\sim 80$\%, since
the FORCE estimates are not numerically accurate when the 
errors are beyond this range. 
}
\label{fig:sn}
\end{figure*}       

Figure~\ref{fig:sn} shows the main results of this paper, the
measurement of the angular amplitudes $s_3$ (triangles) and $s_4$
(squares) in the early SDSS data in four apparent magnitude bins.  The
error estimates were made using the FORCE method, as described in
Section 3.  The open symbols show the results using all the masks of
Scranton \etal (2001); as the Figure shows and as described in Section
3, these measurements cannot be extended to cells much larger than
$0.1^\circ$.  The solid symbols show the results when only the large
masks are used; on scales smaller than $0.1^\circ$, these estimates
are essentially identical to those obtained with all the masks.  As
argued in Section 3, this agreement indicates that the measurements on
larger angular scales using only the large masks should be reliable;
as a result, the measurements can be extended to scales as large as
$\simeq 1.5^\circ$, about half the width of the data stripe.  The
exception to this argument is the measurement of $s_3$ in the faintest
bin, $21 < r < 22$: in this case, there appears to be a
systematic discrepancy between the all-mask and large-mask
measurements even on small scales. We have checked that this is
not due to spurious objects which entered the catalog through
use of the partial masks. All calculations were repeated with
a catalog which was first filtered through the full bright
and faint mask, respectively:
partial masks on the filtered and unfiltered catalogs produce
identical results.

\begin{figure*} 
[htb]
\centerline{\hbox{\epsfig{figure=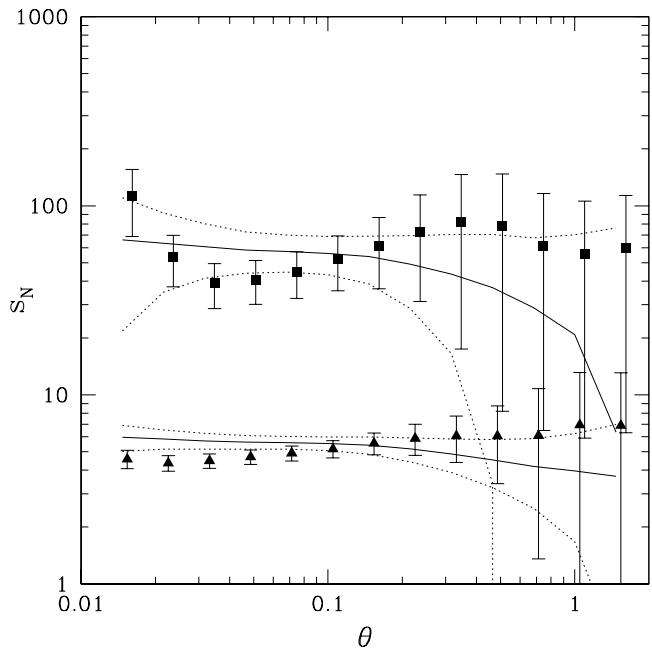,width=9cm}}}
\caption{Results for $s_3$ (triangles) and $s_4$ (squares) for 
$18 < r < 19$, with FORCE errorbars. The solid curves 
show the mean results from 100 {\tt PTHalo} simulations for 
$\Lambda$CDM, with the same survey geometry and selection function 
as the SDSS data. The dotted curves indicate the  
$1-\sigma$ deviations among the simulations.}
\label{fig:force}
\end{figure*}       

To check the accuracy of the FORCE error estimates (in the presence of
the complex masked geometry) and to compare the SDSS results with the
predictions of structure formation models, we show in Figure~\ref{fig:force}
the results for $s_3$ and $s_4$ for the brightest flux slice,
$18 < r < 19$, in eleven logarithmically spaced angular bins in
the range $0.01468^\circ-1.6813^\circ$ (filled symbols, using the
large masks).  We also show the predictions based on 100 mock catalogs
with a radial selection function chosen to match that of the data 
(Dodelson \etal 2001), based on the {\tt PTHalos}
code for a $\Lambda$CDM model (with $\sigma_{8}=0.84$,
$\Omega_{m}=0.3$, $\Omega_{\Lambda}=0.7$, $\Gamma=0.21$ at $z=0$)
with `galaxies' generated as briefly described in Section~3 (see
Scranton \etal 2001 for details).  
The parameters
of this model were chosen in order to reproduce the angular two-point
function $w(\theta)$ measured in the same SDSS data by Scranton \etal
(2001) and Connolly \etal (2001).
The solid curves show the mean of 
the 100 realizations, and the dotted curves show the $\pm 1\sigma$
deviations.  The agreement between the variance estimates on $s_3$ and
$s_4$ from FORCE and {\tt PTHalos} is evident on scales $0.03^\circ <
\theta < 0.3^\circ$.  As noted above, on scales larger than $\sim
0.3^\circ$, where the relative errors are of order 100\% of the
value, or larger,
the FORCE error estimates are not expected to be numerically accurate. 
On scales smaller than $0.03^\circ$, the mock catalogs predict
somewhat higher variance than the FORCE. In any case, the overall
agreement between these error estimates indicates the robustness of
the results.

Figure~\ref{fig:force} also indicates that the measured $s_3$ and
$s_4$ amplitudes are in reasonably good agreement with predictions for
galaxies in the biased $\Lambda$CDM model described above.    
To be more specific, we computed
the goodness of fit of the $\Lambda$CDM model to the data 
using the full covariance matrix from the simulations; 
the resulting reduced
$\chi^{2}$ values are given in the second and third columns of
Table~\ref{Fits}.  Given that this model was not explicitly
constructed to give a good fit to $s_{3}$ and $s_{4}$, it is in
very good agreement with the data.
We verified using mock catalogs that the Gaussian assumption
for the likelihood function is a very good approximation
on small scales, where most of the information is coming from.

In addition, we performed a fit of the SDSS results 
to a hierarchical model (HM), in which $s_3$ and $s_4$ are 
constrained to be constants, independent of cell size, again 
using the covariance matrix\footnote{Typically the cross-correlation 
coefficient between neighboring bins is larger than 0.9, 
and drops to 0.5 for angular scales separated by a factor of 4.}
from the {\tt PTHalos} simulations 
\footnote{For the faint magnitude slices, $r =20-21, 21-22$, for 
which  no explicit covariance matrix from {\tt PTHalos} was available due to
CPU limitations, we used the cross-correlation coefficient
($r_{ij}\equiv C_{ij}/\sqrt{C_{ii}C_{jj}}$) from the $r =19-20$ 
catalogs and scaled up
this matrix by the ratio of the diagonal ($C_{ii}$) 
FORCE errors for the different simulation slices. This ansatz was 
tested by using it to scale from the $18-19$ to the $19-20$ slices, in which  
case it accurately reproduced the measured covariance matrix for the 
fainter slice. Here $C_{ij}$ is the covariance matrix of the $s_p$ 
estimator for cell sizes $\theta_i$ and $\theta_j$.}.   
The resulting higher order 
amplitudes and $\chi^2$ values are 
shown in the fourth through seventh columns of
Table~\ref{Fits}.  The relatively low reduced $\chi^2$ values 
indicate that the resulting constant amplitudes provide a very
good description of the data given the errors.
Although there are apparent deviations from exact 
hierarchical scaling in Fig.~\ref{fig:sn}, these are not statistically  
significant.

\begin{deluxetable}{ccccccc}
\tablewidth{14cm} \tablecaption{The first two columns give the
$\chi^{2}$ per degree of freedom for the comparison of the SDSS data with 
the $\Lambda$CDM model.  The remaining columns give the
best-fit constant values for $s_3$ and $s_4$ (hierarchical model, HM) 
and their goodness of fit. 
All fits use the 
covariance of the measurements between different angular scales estimated 
from the mock catalogs.  }
\tablehead{\colhead{$r$} & \colhead{$\chi^{2}_{\rm
\Lambda CDM}(s_{3})$} & \colhead{$\chi^{2}_{\rm \Lambda CDM}(s_{4})$} &
\colhead{$s_{3}^{\rm HM}$} & \colhead{$\chi^{2}_{\rm HM}(s_{3})$} &
\colhead{$s_{4}^{\rm HM}$} & \colhead{$\chi^{2}_{\rm HM}(s_{4})$} }
\startdata 18-19 & 1.4 & 1.4 & $4.9 \pm 0.4$ & 0.7 & $42\pm 11$ & 1.2
\nl 19-20 & 1.8 & 1.6 & $4.1 \pm 0.4$ & 0.8 & $38\pm 9 $ & 1.1 \nl
20-21 & --- & --- & $4.2 \pm 0.4$ & 0.7 & $41\pm 10$ & 0.7 \nl 21-22 &
--- & --- & $3.6 \pm 0.3$ & 0.8 & $41\pm 10$ & 0.7 \nl \enddata
\label{Fits}
\end{deluxetable}

\begin{figure*} 
[htb]
\centerline{\hbox{\epsfig{figure=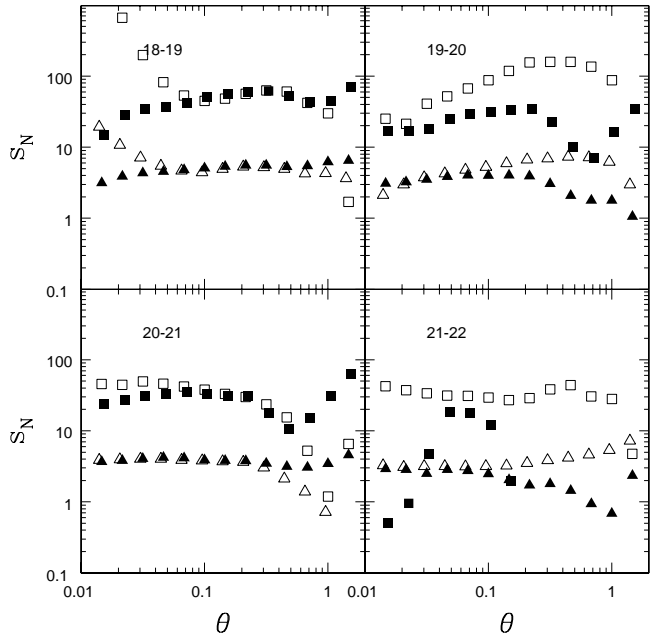,width=9cm}}}
\caption{$s_3$ and $s_4$ measurements for ``elliptical'' (open symbols)  
and ``spiral'' galaxies (closed symbols) in early SDSS data. Only 
measurements with the large masks are shown, and errorbars are 
omitted for clarity. Objects in this sample were selected to have 
$p_{gal} \ge 0.99$ according to the Bayesian star-galaxy separator (see 
Sec. 4.2).}  
\label{fig:snES}
\end{figure*}

The high-quality CCD imaging in 5 passbands enables the SDSS to 
classify galaxies based on their photometric properties and therefore 
allows investigation of the clustering of galaxies as a function of 
galaxy type. As a first exercise at this, we have crudely divided the 
sample described above by morphological type: ``ellipticals'' 
designate objects for which the de Vaucouleurs profile fit is of 
higher likelihood than the exponential profile according to the 
{\tt photo} pipeline, and ``spirals'' constitute the remainder.
We have made no attempt to assess the reliability of this classification 
scheme as a function of magnitude 
or to correlate it with other photometric properties. 
The resulting $s_p$ measurements for ellipticals (open symbols) and 
spirals (closed symbols) are shown in Figure~\ref{fig:snES}. 
These results are suggestive: for example, the tendency 
for higher clustering amplitudes for the ellipticals, particularly 
on small scales, is at least qualitatively 
consistent with the well-known morphology-density relation. However, 
there is no clearly discernible trend in the results 
from bright to faint magnitudes, and 
the differences are for the most part not statistically significant given 
the errors on this relatively small sample. While we cannot draw 
firm conclusions from this first assay, it is evident that the full 
SDSS dataset should provide an excellent sample for studying higher 
order clustering as a function of galaxy type, especially if
colors and spectral types are used in the selection (e.g., Ivecic \etal
2001).

\subsection{Auxiliary Measurements: Checks for Systematic Errors}

We now describe a number of tests carried out in order to gauge 
possible systematic effects on the results. While the tests described 
by Scranton \etal (2000), although performed for two-point
statistics, should imply the reliability of  
higher order statistics measurements, some extra checks are warranted.

\begin{figure*} 
[htb]
\centerline{\hbox{\epsfig{figure=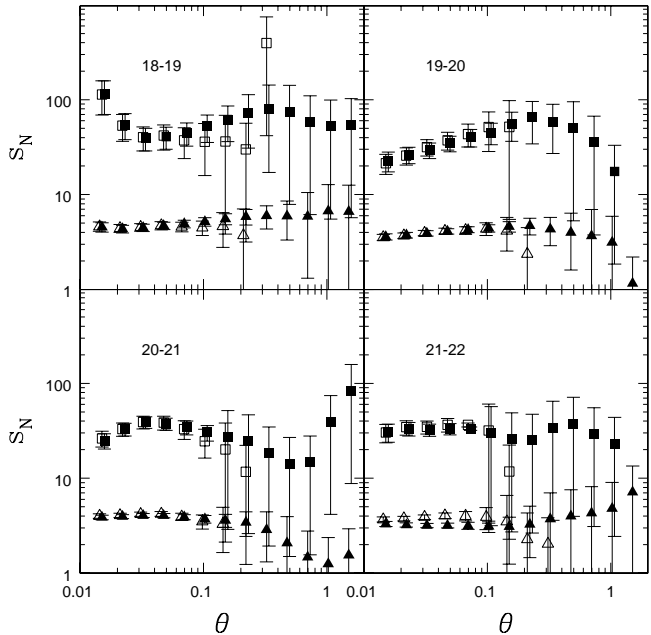,width=9cm}}}
\caption{Same as Figure~\ref{fig:sn}, but for a catalog containing 
galaxies selected with Bayesian probability $p_{gal} \ge 0.99$.}
\label{fig:snp}
\end{figure*}

In the results shown above, we used the SDSS image processing 
pipeline to perform star-galaxy separation: the resulting 
catalog contained all objects selected as galaxies by {\tt photo}. 
As noted in Section 2, the Bayesian star-galaxy separation 
described in Scranton \etal (2001) provides a more accurate 
method at faint magnitudes, by assigning stellar and galactic 
probabilities to each object. 
To test the effects of possible stellar contamination of the 
{\tt photo}-selected sample,
we have repeated the $s_p$ measurements using the Bayesian
probabilities, by constructing a catalog containing all objects with 
galactic probability $p_{gal} \ge 0.99$. The results, shown in 
Figure~\ref{fig:snp}, 
are almost indistinguishable from the {\tt photo}-selected 
results of Figure~\ref{fig:sn}. In the  
faintest magnitude slice, there appears to be a small difference 
in the results, but it is not statistically significant.
This test indicates that stellar contamination of the sample 
is not a significant source of systematic error in the measurement 
of $s_3$ and $s_4$. 

\begin{figure*} 
[htb]
\centerline{\hbox{\epsfig{figure=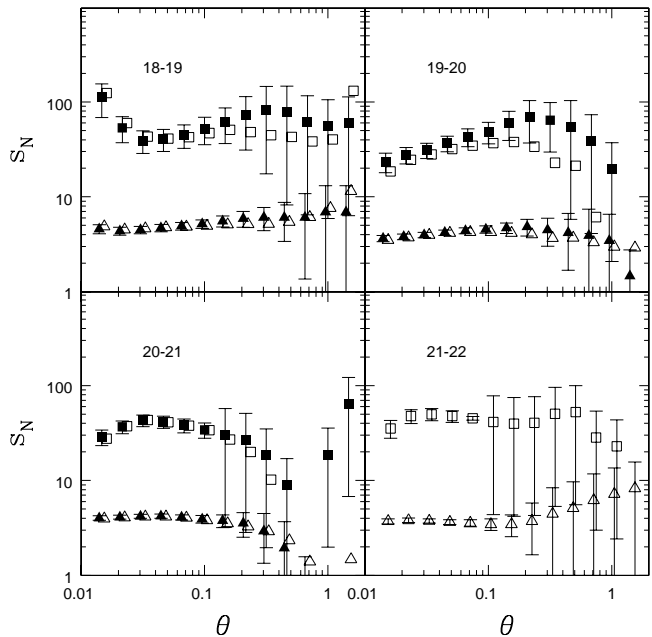,width=9cm}}}
\caption{$s_3$ and $s_4$ measured with the Scranton \etal (2001) masks
(solid symbols) compared with results using the more stringent
$1.6^{\prime\prime}$ `faint' seeing mask for the three brightest flux
slices (open symbols).  The differences are well within the
statistical errors.  For clarity, errorbars are only plotted for the
principal measurements.}
\label{fig:sns}
\end{figure*}

Another potential source of systematic error is incomplete masking of
areas of poor-quality data, particularly bad seeing.  The results
shown above used the masks recommended by Scranton \etal (2001): the
`bright' mask (used for $18 < r < 21$) excludes regions with
seeing $> 1.75^{\prime\prime}$, and the more conservative `faint' mask
(used for $21 < r < 22$) excludes regions where the seeing $>
1.6^{\prime\prime}$.  Scranton \etal (2001) showed that these cuts
yield a galaxy dataset with negligible cross-correlations with
systematic effects such as stellar density, seeing, and dust
extinction.  As a further test, we have repeated the $s_p$ estimates
by excluding data with seeing $>1.6^{\prime\prime}$ in all four
magnitude bins, i.e., applying the more stringent `faint' seeing mask
to the brighter ($18 < r < 21$) data as well.  The results are
shown in Figure~\ref{fig:sns}: solid symbols indicate the `bright'
seeing mask of $1.75^{\prime\prime}$ and open symbols indicate results
for the more conservative `faint' seeing mask of $1.6^{\prime\prime}$. 
In both cases, we show results using only the large masks.  
There is evidence for small systematic shifts in the $s_p$ amplitudes
between the two masks, particularly for the two brightest slices; 
we speculate that the reduced effective survey area for the data with 
the `faint' mask could lead to a small cosmic bias (a systematic 
underestimate) in the estimate of the $s_p$. However, the more important 
conclusion from this comparison is that 
almost all the changes are well within the 1-$\sigma$ 
errors, indicating 
that systematic errors due to incomplete masking are well
within the statistical errorbars.  (Note that the $s_p$ estimates at
different $\theta$ are correlated, so they cannot simply be combined
to increase the significance of any differences.)  To further test the
effect of masks, we also computed the higher-order moments without
using any mask at all.  As shown in Figure~\ref{fig:snapm}, this
results in an underestimate of $s_3$ and $s_4$ comparable to the
differences seen in Figure~\ref{fig:sns}.  This should represent a
conservative upper limit on possible systematic errors due to
incorrect masking.

As a further check, we have used the 100 {\tt PTHalos} simulations of
the $18 < r < 19$ and $19 < r < 20$ magnitude slices to
check if the masks introduce any systematic errors into our
measurements.  We have performed three measurements in each mock SDSS
catalog: (i) with no mask, (ii) with the `bright' (large) masks, and
(iii) with the `faint' (large) masks, using the same CIC method as
applied to the data.  As expected, the errorbars were larger for the
samples with more masked regions (smaller effective survey area), with
the effect being more significant on large angular scales.  On most
scales, there is no systematic shift in the mean values of the $s_p$
(no bias) between the 3 cases, although there is a slight bias toward
lower $s_p$ values on the very largest scales for the samples with 
more masked regions. As speculated above, it is possible that an
effective cosmic bias due to the skewness of the distributions of 
$s_3$ and $s_4$ 
values leads to a systematic underestimate of these amplitudes
at large scales when the masked area is increased. 
We have attempted to quantify this effect from the 
mock catalogs, but the relatively small number of realizations (100) 
precludes us from reaching a definite conclusion. In any case, 
the amplitude of the bias of the $s_p$ estimates is
negligible compared to the measured variance.  We conclude that no
appreciable bias in the estimates of $s_3$ and $s_4$ can be attributed
to the complicated geometry of the survey region.

\section{Comparison with Previous Measurements}

It is instructive to compare the early SDSS results for 
$s_3$ and $s_4$ with previous measurements. To compare 
with results from the APM and  
EDSGC surveys as well as with the early SDSS results reported 
by Gazta\~naga (2001a), we have measured CIC statistics 
for SDSS galaxies in the 752/756 dataset 
with $15.9 < r < 18.9$. This choice is based on that of 
Gazta\~naga (2001a), who demonstrated that 
the apparent magnitude range $15.9 < g < 18.9$ in the SDSS  
yields a galaxy surface density comparable to the 
$b_J < 20$ samples of the APM and EDSGC surveys. The $r$-selected 
sample used here will have a slightly different effective depth than his 
$g$-selected sample, but $s_3$ and $s_4$ should be relatively 
insensitive to this. A more substantive difference likely arises 
from the fact that samples selected in different passbands will contain 
different mixes of galaxy types and therefore yield different clustering 
amplitudes (see Figure~\ref{fig:snES}); as a result, we do not 
expect these samples to yield identical higher order moments, but they 
should be qualitatively similar. 

\begin{figure*} 
[htb]
\centerline{\hbox{\epsfig{figure=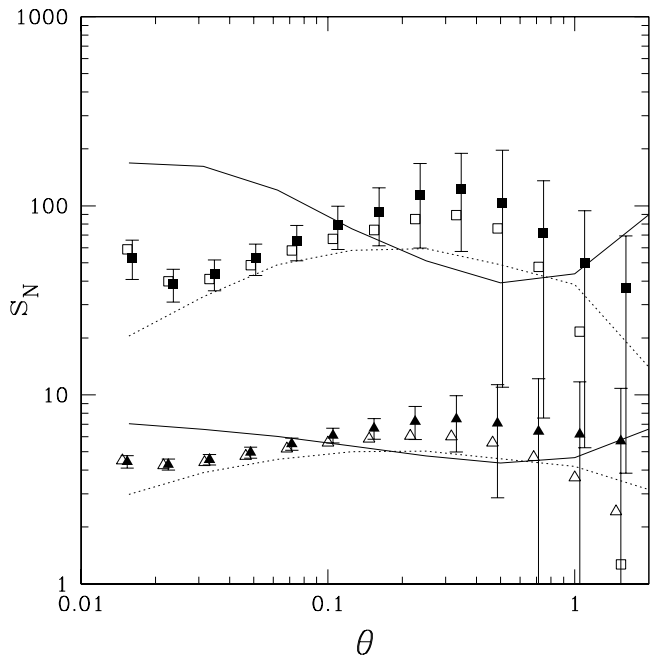,width=9cm}}}
\caption{$s_3$ and $s_4$ measurements in the SDSS for 
$15.9 < r < 18.9$ (closed symbols with FORCE errorbars) compared with 
results from the APM (dotted curves) and EDSGC (solid curves) surveys.
Open symbols without error bars show the SDSS results when the 
masks are not used.}

\label{fig:snapm}
\end{figure*}       

Figure~\ref{fig:snapm} shows the 
results of this comparison for $s_3$ (triangles and lower curves) and 
$s_4$ (squares and upper curves) for the 
SDSS data with large masks (closed symbols with errors), the EDSGC 
measurement (solid curves, Szapudi, Meiksin, \& Nichol 1996), 
and the APM survey (dotted curves, Gazta\~naga 1994, Szapudi \etal 1995, 
Szapudi \& Gazta\~naga 1999). All measurements displayed here used 
the same CIC algorithm. The APM and EDSGC errorbars are omitted for 
clarity, since they are smaller than those for this bright SDSS sample 
which covers less area. On intermediate to large scales, 
the SDSS results are 
consistent at the 1.5 $\sigma$ level with both the APM and the
EDSGC measurements.  Going to small scales, the $s_3$ amplitude in the
SDSS appears to be intermediate between the EDSGC and APM values 
and remains 
fairly flat, while the latter two are rising and falling, respectively.
This result is not too surprising: since the 
EDSGC and APM catalogs 
are based on digitized maps 
of essentially the same photographic plates, the 
differences between 
their results on small scales must be largely due to systematic 
effects in one or both of these catalogs. As noted by Szapudi \&
Gazta\~naga 1999, the APM (EDSGC) amplitudes are likely biased low  
(high) by their different methods of deblending objects in crowded fields. 
For the kurtosis, $s_4$, 
the SDSS results on small scales are also intermediate between 
the other two surveys, although somewhat closer to the APM values.
Overall, the consistency between the measurements of the higher order 
moments in these different datasets is encouraging.

The results for $s_3$ in the SDSS shown in Figure~\ref{fig:snapm} 
appear to be in qualitative agreement with 
those obtained by Gazta\~naga (2001a) on scales smaller than $0.1^\circ$, 
but they are systematically higher than his results on larger scales. 
There are several possible reasons for this discrepancy, and they 
may act in combination. The data in the survey region used in his analysis 
(Runs 94/125, Southern Equatorial Stripe), 
which does not overlap the region analyzed here, 
was taken for the most part in relatively
poor seeing conditions early in the 
commissioning phase of the project.
Application of the `bright' seeing cut ($1.75^{\prime\prime}$) recommended 
by Scranton \etal (2001) for the removal of systematic effects 
would mask out roughly 37\% of the 94/125 area. Along with use 
of an undersampled CIC method, not including the masks 
could lead to an underestimate  
of $s_3$ (Szapudi, Meiksin, \& Nichol 1996; Szapudi \& Gazta\~naga 1998). 
As an example, the open symbols in 
Figure~\ref{fig:snapm} show the results for $s_3$ 
and $s_4$ for the 752/756 dataset when the `bright' mask, which covers 
about 20 \% of this data region, is not used. In addition to the 
mask, cosmic variance could also contribute to the difference in 
$s_3$ results seen on large scales. 
Other small differences 
in the data selection between Gazta\~naga 2001a and this paper 
($g$ vs. $r$ selection, use of uncorrected vs. 
extinction-corrected magnitudes, use of Petrosian vs. model magnitudes) 
are probably less significant and unlikely to account for the 
difference in results. In fact, as we were about to submit this paper 
for publication, Gazta\~naga 2001b appeared, which also included an 
analysis of $s_3$ for the EDR north slice which we analyze in this paper. 
His results for this data slice are remarkably consistent with the open 
triangles in Figure~\ref{fig:snapm}, in which the mask is not used. 

\begin{figure*} 
[htb]
\centerline{\hbox{\epsfig{figure=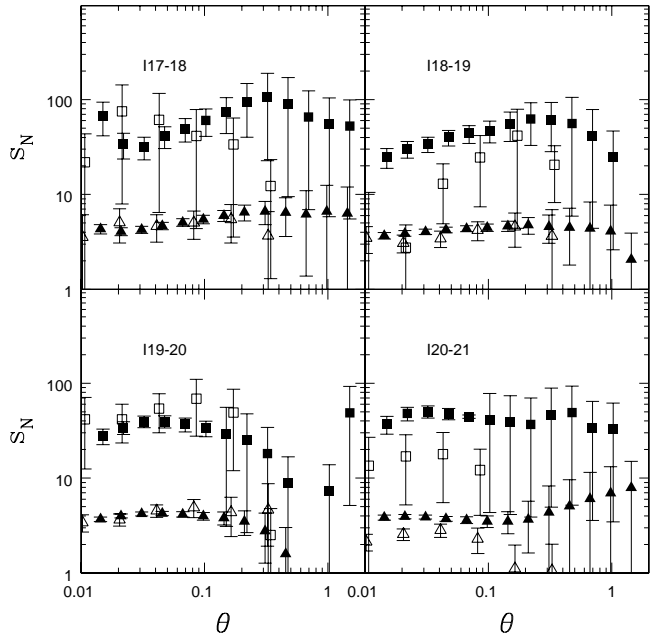,width=9cm}}}
\caption{$s_3$ and $s_4$ for the Deeprange survey (open symbols), 
in $I$ band magnitude slices $17-18, ..., 20-21$, compared with 
SDSS results in the approximate corresponding $r$ slices 
$17.9-18.9, ..., 20.9-21.9$ (solid symbols).
For each survey, the appropriate large masks were used,  
and the errorbars were calculated 
by the FORCE package.}
\label{fig:sndr}
\end{figure*}

We turn next to comparison of the SDSS results with 
measurements in the Deeprange survey (Szapudi \etal 2000). The 
Deeprange catalog was selected in $I$; for a typical galaxy SED, 
and assuming a mixture of early and late types,
the approximate correspondence between SDSS $r$ and $I$ band is 
$r_{sdss} \simeq I_{Deeprange} + 0.9$
(Postman 2001, private communication).  We therefore 
constructed new SDSS samples in the $r$ 
ranges $17.9-18.9, 18.9-19.9, ...$ to compare with 
the Deeprange samples of $I=17-18, 18-19, ...$. These SDSS samples are 
shifted by just 0.1 mag from those analyzed in Sec. 4.  
We note that 
the Deeprange magnitudes were not corrected for extinction, which 
should be negligible over this small, high-latitude field.
The results of this comparison for $s_3$ and $s_4$  
are shown in Figure~\ref{fig:sndr}, with solid symbols corresponding 
to the SDSS data and open symbols to the Deeprange.  The figure panels are 
labeled by the Deeprange $I$-band range, e.g., I$19-20$ is compared with 
$r = 19.9-20.9$ in the SDSS.

As expected, comparison of Figure~\ref{fig:sndr} with 
Figure~\ref{fig:sn} shows 
that the $0.1$ magnitude shift in the SDSS magnitude ranges causes very 
little change in the measured $s_p$'s.  The SDSS results for $s_3$ are in
excellent agreement with the Deeprange values in the three
brightest slices, while the values are $\simeq 1-2\sigma$ discrepant 
in the faintest SDSS slice. This suggests that the early SDSS data  
represents the most accurate measurement to date 
of the angular skewness for magnitudes $r < 21$.  
For $s_4$, there is also qualitatively good agreement between 
the two surveys for the 3 brighter magnitude slices, with a few discrepant 
points; this is not surprising, since the kurtosis is more sensitive 
to both systematic errors and cosmic variance than the skewness.

The comparison for the faintest SDSS slice $r = 21-22$ is
qualitatively different from the others: while the Deeprange data 
indicate a systematic falling off of the amplitudes with increasing 
sample depth, this trend is either absent or 
not significant in the SDSS data (see also Table~\ref{Fits}).
Given possible differences due to cosmic variance or 
undetected systematic errors, the current data cannot decide between 
these two trends. The answer, which has important implications for the 
evolution of galaxy bias, should come when a larger SDSS sample is 
available. 

\section{Conclusion}

We have measured the moments of angular counts in cells (CIC)
in early SDSS data, using the Northern Equatorial Stripe. 
Although the results presented here focus on the higher order 
moments, we have also checked 
that the measured second order moments from CIC 
are consistent with the direct measurements of the angular two-point
correlation function presented in Connolly \etal (2001) and Scranton 
\etal (2001).  To control
statistical and systematic errors, we have used state of the art
measurement techniques and the highest quality segment of the
available data.  This resulted in a set of complicated masks, which
probably does influence the accuracy of our measurement via edge
effects.  We have shown with auxiliary measurements and simulations
that this should not produce any significant bias, only extra variance.
According to Figure~\ref{fig:snp}, star-galaxy 
separation has insignificant contaminating effects on the $s_p$
measurements, except perhaps for the faintest magnitude slice.

Figures~\ref{fig:sn}-\ref{fig:force} and Table~\ref{Fits} 
show that the $s_p$'s behave qualitatively 
as expected from theories of hierarchical
structure formation.  The $s_3$ and $s_4$ amplitudes display an 
approximate hierarchical scaling,
with deviations consistent with cosmic (co)variance.  As quantified in
Table~\ref{Fits}, the measurements of $s_3$ and $s_4$ are consistent
with the predictions of non-linear 
$\Lambda$CDM models, in which the galaxy bias is determined 
by the fact that the number
of galaxies in halos of mass $m$ scales as $N_{gal} \propto m^{0.8-0.9}$. 
While this conclusion should be regarded as preliminary,
given the limitations of the current data set and the restricted
number of models considered, it is worth noting that a similar model 
matches the APM power spectrum and $s_p$'s on small scales
(Scoccimarro \etal 2001).

We have also presented preliminary results for $s_3$ and $s_4$ for 
galaxies separated by morphological type into `ellipticals' and `spirals', 
based on profile fits to the images by the {\tt photo} pipeline.
As expected from the morphology-density relation, it 
appears that `ellipticals' are more strongly clustered than `spirals'; 
the robustness of this result remains to be demonstrated with a larger 
sample and a more refined method of morphological classification.

The Sloan Digital Sky Survey (SDSS) is a joint project of The
University of Chicago, Fermilab, the Institute for Advanced Study, the
Japan Participation Group, The Johns Hopkins University, the
Max-Planck-Institute for Astronomy (MPIA), the Max-Planck-Institute
for Astrophysics (MPA), New Mexico State University, Princeton
University, the United States Naval Observatory, and the University of
Washington.  Apache Point Observatory, site of the SDSS telescopes, is
operated by the Astrophysical Research Consortium (ARC).

Funding for the project has been provided by the Alfred P. Sloan
Foundation, the SDSS member institutions, the National Aeronautics and
Space Administration, the National Science Foundation, the U.S.
Department of Energy, the Japanese Monbukagakusho, and the Max Planck
Society.  The SDSS Web site is http://www.sdss.org/.  IS thanks
Stephane Colombi for useful discussions and  was partially supported
by a NASA AISR, NAG-10750.  We acknowledge the use of the FORCE
(FORtran for Cosmic Errors) package by S. Colombi and IS, 
available from http://www.ifa.hawaii.edu/$^{\sim}$szapudi/istvan.html.

\end{document}